\documentclass{ptapap}

\author{Greg Stachowski}[UP]
\affil[UP]{Mt. Suhora Observatory, Pedagogical University of Cracow,
Podchorążych 2, 30-084 Krakow, Poland}
\author{Tomasz Kundera}[UJ]
\affil[UJ]{Jagiellonian University Astronomical Observatory, Orla 171, 30-244 Krakow, Poland}
\author{Paweł Ciecielag}[CAMK]
\affil[CAMK]{N. Copernicus Astronomical Center, Bartycka 18,
00-716 Warszawa, Poland}
\author{the AstroGridPL team}

\title{AstroGrid-PL}

\begin{document}
\setcounter{page}{292}

\maketitle

\begin{abstract}
We summarise the achievements of the AstroGrid-PL project, which aims to
provide an infrastructure grid computing, distributed storage and 
Virtual Observatory services to the Polish astronomical community.
It was developed from 2011--2015 as a \textit{domain grid} component
within the larger PLGrid Plus project for scientific computing in Poland. 
\end{abstract}

\section{Introduction}

Improvements in detector technology mean that today astronomy, like 
other experimental sciences, is dominated by
the rapid accumulation of extremely large data sets. For example, a typical 
single night of observations at a telescope equipped with a CCD camera can produce several gigabytes of raw data, while new survey telescopes such as the LSST are projected to produce up to 1 terabyte of data per night. This means that it is no longer possible for a single researcher to store and process the data on their own workstation, and indeed it is not even possible for a single researcher to 
analyze all the available data contained in even their own observations (which may contain information about hundreds or thousands of additional objects in the same field of view as the target star or galaxy). Thus a new approach to data storage, retrieval and processing is required, which harnesses geographically distributed computing resources to provide storage, processing, and sharing of these large data sets in a manner transparent to the end user. These new approaches are variously known 
as the \textit{cloud}, the \textit{grid} and, in an astronomical context, the \textit{virtual observatory}.

\section{Aims of the AstroGrid-PL project}

The idea of what became the AstroGrid-PL project first arose in around 2009, with the initial aim
of building data centres at several major astronomical institutions in Poland and using them to
provide the services described above to the Polish astronomical community. Ultimately, however, AstroGrid-PL was funded not as an independent project but as part of the second phase of development of a general computing infrastructure for science in Poland, known as PLGrid. This second phase, PLGrid Plus, was focussed on \textit{domain grids}, each providing software development, integration and user support dedicated to specific fields of scientific research (e.g. quantum chemistry, bioinformatics, astronomy, high-energy physics etc.) based on hardware provided by the Polish supercomputing centres and developed during Phase 1 (e.g. ACK Cyfronet in Krakow, PCSS in Poznan, etc.) \citep{Kitowski2014}.
 
The aims of AstroGrid-PL specifically were to provide \textit{(a)} a distributed storage system for astronomical data sets with metadata and automated replication, \textit{(b)} simplified access for users to grid computing resources via graphical workflows, \textit{(c)} a virtual observatory for access and sharing of astronomical data and to interoperate with and teach users about the international virtual observatory system, and \textit{(d)} to digitise archival data, particularly photographic plates, which were in storage at various observatories in Poland and were in danger of loss due to age and accidental damage. Secondary aims were to use the project to more closely integrate the Polish astronomical community and foster cooperation, through provision of videoconferencing services, a web portal for services, and workshops and training. Access to the resources is available free of charge to any scientist with an entry in the OPI database of Polish scientists, or their students and co-workers.

\section{Results}

The aim of a distributed data storage system, ``Astro-data'', was achieved using the iRods distributed storage software. This provides distributed storage (at the supercomputing data centres) with automatic replication, text-based and graphical user interfaces (developed as part of the project), fine-grained access control including  embargoing of data before a certain date, and ``tickets'' for easy sharing of data. This system is now available for the Polish astronomical community \citep{2014LNCS.8500..305K}.

\begin{figure}
\centering
\includegraphics[width=0.9\textwidth]{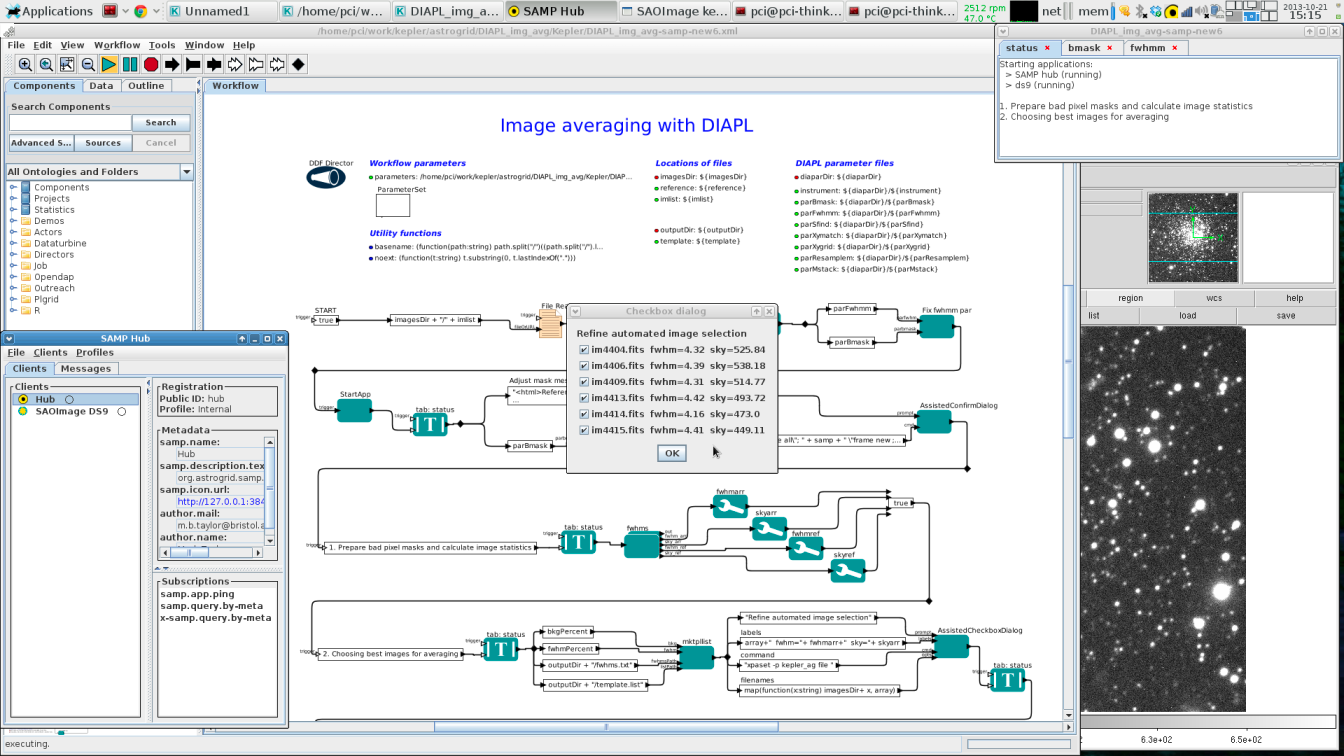}
\caption{The DIAPL difference image analysis workflow in the \textit{Kepler} environment.}
\label{fig:bb}
\end{figure}

Access to grid computing resources is provided firstly by ``Astro-pipelines'' which uses the \textit{Kepler} and \textit{Reflex} workflow environments to allow processing of data (stored within the ``Astro-data'' iRods system) through a graphical workflow environment. Example workflows developed during the project include image subtraction (DIAPL, \ref{fig:bb}) and calculation of broadening functions for spectroscopic data \citep{Ciecielag2014}. Secondly, astrophysical fluid dynamics simulations have been developed using the InSilicoLab framework, again providing a simplified graphical environment for running simulations with various parameters and initial conditions using the grid infrastructure \citep{Hanasz2014}.

\begin{figure}
\centering
\includegraphics[width=0.9\textwidth]{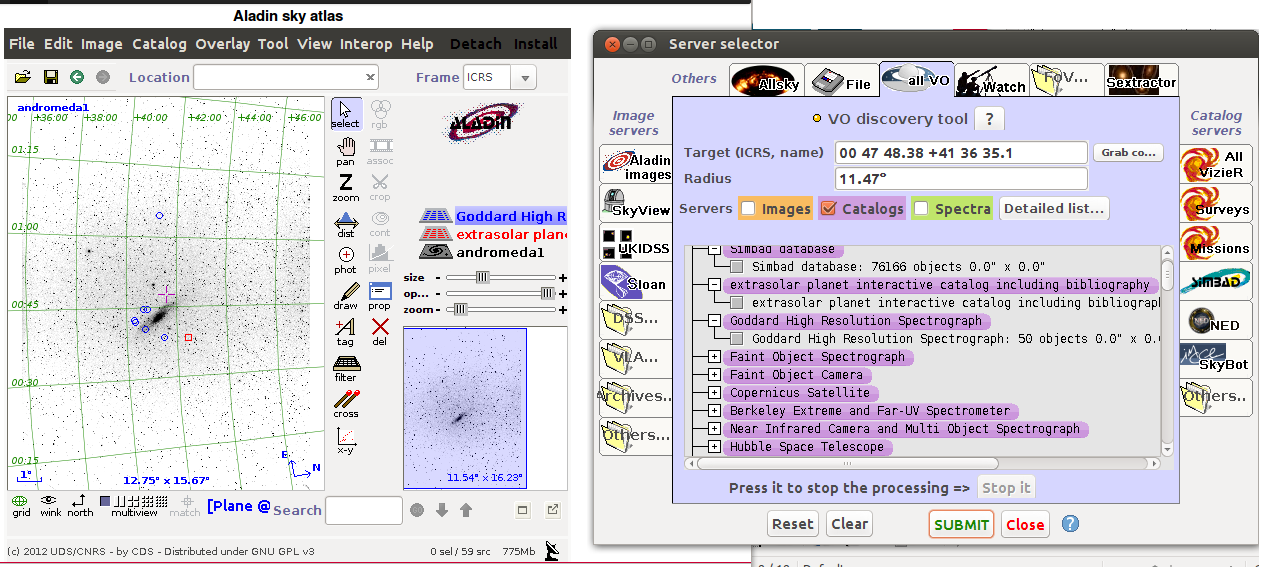}
\caption{Image of digitised archival photographic plate imported into Aladin with overlaid coordinate grid (determined as part of the digitisation process) and positions of known objects from selected catalogues.}
\label{fig:aa}
\end{figure}

The Polish Virtual Observatory \citep{2014LNCS.8500..305K} provides access to archival data sets using standard virtual observatory protocols, thus allowing existing software such as IRAF, Topcat or Aladin (see fig.~\ref{fig:aa}) interact seamlessly with existing data. The Virtual Observatory has been initially populated with a selection of digitised photographic plates converted to FITS format. Several thousand plates have been scanned using a dedicated large-format scanner bought specifically for the project \citep{2014aspl.conf...48K}. These will also be made available as the slow process of entering metadata (times, dates, observer information, etc.) from handwritten log books is completed.

A number of workshops and training sessions have been held, and the videoconferencing systems put in place for the project are regularly used to transmit seminars and lectures between the participating institutions.  

Finally the AstroGrid-PL portal\footnote{\texttt{https://astrogrid-pl.org/home}} provides a single point of access to the AstroGrid-PL services and documentation and registration with PLGrid Plus.

\acknowledgements{We would like to thank the Polish astronomical institutions participating in AstroGrid-PL, and particularly the N. Copernicus Astronomical Center of the Polish Academy of Sciences.}

\bibliographystyle{ptapap}
\bibliography{astrogrid}

\end{document}